\documentclass[twocolumn,showpacs,superscriptaddress]{revtex4}%
\usepackage{amsfonts}
\usepackage{amsmath}
\usepackage{amssymb}
\usepackage{graphicx}%
\usepackage{gensymb}
\usepackage{xcolor}
\providecommand{\U}[1]{\protect\rule{.1in}{.1in}}

\begin{document}
\title{Evolution of the topologically protected surface states in superconductor $\beta$-Bi$_{2}$Pd from the three-dimensional to the two-dimensional limit}
\author{Bao-Tian Wang}
\affiliation{Department of Physics, Applied Physics and Astronomy,
Binghamton University-SUNY, Binghamton, NY 13902, USA}
\affiliation{Institute of High Energy Physics, Chinese Academy of
Sciences (CAS), Beijing 100049, China} \affiliation{Dongguan Neutron
Science Center, Dongguan 523803, China}
\author{Elena R. Margine}
\affiliation{Department of Physics, Applied Physics and Astronomy,
Binghamton University-SUNY, Binghamton, NY 13902, USA}


\begin{abstract}
{The recent discovery of topologically protected surface states in
the noncentrosymmetric $\alpha$-BiPd and the centrosymmetric
$\beta$-Bi$_{2}$Pd has renewed the interest in the Bi-Pd family of
superconductors. Here, we employ first-principles calculations to
investigate the structure, electronic, and topological features of
$\beta$-Bi$_{2}$Pd, in bulk and in thin films of various
thicknesses. We find that the Van der Waals dispersion corrections
are important for reproducing the experimental structural
parameters, while the spin-orbit interaction is critical for
properly describing the appearance of topological electronic states.
By increasing the thickness of the slab, the Dirac-cone surface
states and the Rashba-type surface states gradually emerge at 9 and
11 triple-layers.}

\end{abstract}
\maketitle

\section{Introduction}

The prediction and discovery of topological insulators
(TIs)~\cite{Fu07,Hsieh08,ZhangHJ,XiaY,Hsieh09,Chen09,Liu,Hasan10,Sawai10,Eremeev10,Yang,Kolmogorov16,Bansil16}
have triggered the quest for other classes of materials that host exotic quantum states,
such as topological superconductors~\cite{XLQi2,Hasan15,Sato16} and Weyl~\cite{Huang_NCOM15,Weng_PRX15,Lv_NPHYS15,Yang_NPHYS15,Xu_NPHYS15}, Dirac~\cite{Wang_PRB12,Ali14,Borisenko_PRL14,Liu_SCI14,Neupane_NCOM14}, and nodal-line~\cite{Bian_NCOM16,Wu_NPHYS16} semimetals. Similar to TIs, topological superconductors (TSCs) are characterized by a full
paring gap in the bulk and topologically protected gapless states on
the edge or surface that can support massless Majorana
fermions~\cite{XLQi2,Schnyder-PRB15,Ando15,Xu15,Hasan15}. These exotic
states may find promising applications in spintronics and quantum
computing~\cite{Fu08,Nayak}.

\begin{figure*}
\begin{center}
\includegraphics[width=0.8\linewidth]{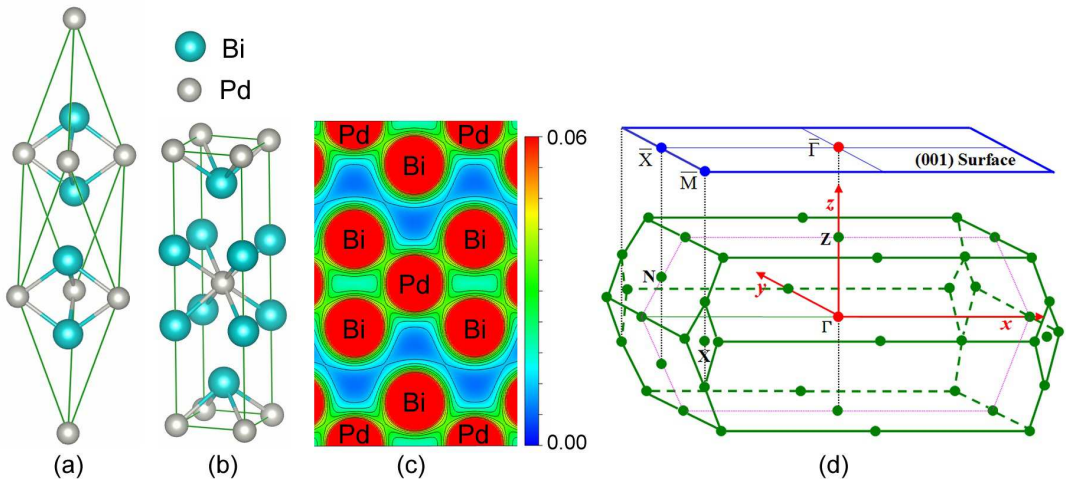}
\end{center}
\caption{(Color online) (a-b) Crystal structure of
$\beta$-Bi$_{2}$Pd in the primitive and conventional body-centered
tetragonal unit cells. (c) Charge density distribution in the (110)
plane of the conventional crystal structure. The contour lines are
drawn from 0.0 to 0.06 at 0.006 $e$/au$^{3}$ intervals. (d) Bulk BZ
for the primitive unit cell of $\beta$-Bi$_{2}$Pd. The square plane
represents the two-dimensional BZ of the projected (001) surface.
}%
\label{fig1}%
\end{figure*}

Although numerous materials have been identified as TIs, only a handful of systems have been
found to exhibit the signatures of topological superconductivity.
The proposed candidate TSCs have been realized by
introducing carriers into topological insulators, such as
Cu-intercalated Bi$_{2}$Se$_{3}$~\cite{Fu10,Hor,Kriener,Sasaki11}
and In-doped SnTe \cite{Sasaki12}, or by applying pressure to
topological parent compounds, such as Bi$_{2}$Te$_{3}$,
Sb$_{2}$Te$_{3}$, and Bi$_{2}$Se$_{3}$
\cite{ZhangPNAS,ZhuSR,Kirshenbaum-PRL,KongPP}. More recently, two
members of the Bi-Pd family of superconductors have been found to
exhibit topologically protected surface states with Rashba-like spin
splitting. While the superconducting state of noncentrosymmetric
BiPd appears to be topologically trivial~\cite{Mondal,Sun,Neupane},
the pairing symmetry of centrosymmetric $\beta$-Bi$_{2}$Pd remains
to be elucidated~\cite{Sakano,Herrera,Che,Kacmarcik,Biswas,Margine,LvYF}.
Despite the fact that spin- and angle-resolved photoemission spectroscopy (ARPES) has revealed the presence of notrivial surfaces states at the Fermi level~\cite{Sakano}, no Andreev bound states associated with Majorana fermions have been observed in point-contact spectroscopy~\cite{Che}. On the other hand, according to a recent scanning tunneling microscopy study, Majorana zero modes have been identified in $\beta$-Bi$_{2}$Pd crystalline films~\cite{LvYF}. Along with the recently discovered PbTaSe$_{2}$ superconductor~\cite{Ali_PRB14,Bian_NCOM16,Chang_PRB16}, $\beta$-Bi$_{2}$Pd constitutes a promising candidate for the long-sought stoichiometric TSC (i.e, a three-dimensional TSC realized in the absence of any external
factors such as doping or pressure).

In the present work, first-principles calculations are carried out
to systematically study the thickness-dependent evolution of the
electronic structure and topological properties in
$\beta$-Bi$_{2}$Pd thin films. The bulk material has a layered
structure consisting of stacked Bi--Pd--Bi triple layers (TLs).
Owing to the weak coupling between two adjacent TLs, it has been
possible to achieve TL-by-TL growth of $\beta$-Bi$_{2}$Pd thin films
using molecular beam epitaxy~\cite{LvYF}. This technique has also
been used to grow high-quality thin films of Bi$_{2}$Se$_{3}$,
Bi$_{2}$Te$_{3}$, and Sb$_{2}$Te$_{3}$ of various thicknesses by
precisely controlling the growth
conditions~\cite{He,Li_AdvMat10,WangG}. In addition to the
electronic properties of thin films, we also investigate the effect
of van der Waals (vdW) and spin-orbit interaction (SOI) on the
structure and the band dispersion of bulk $\beta$-Bi$_{2}$Pd.

\section{Methods}

The calculations are performed at the density functional theory level, employing the
projector augmented wave method as implemented in the VASP package
\cite{Kresse3}. The Perdew, Burke, and Ernzerhof (PBE) \cite{PBE}
form of the generalized gradient approximation is chosen to describe
the exchange-correlation energy. Bi 5$d^{10}$6$s^{2}$6$p^{3}$ and Pd
4$p^{6}$4$d^{9}$5$s^{1}$ orbitals were included as valence
electrons. To properly treat dispersion corrections, we use the
non-local vdW density functional optB86b-vdW~\cite{vdW1,vdW2}.
Relativistic effects are included in the calculations in terms of
the SOI~\cite{SOI}. The energy cutoff of the plane-wave basis-set is
500~eV. The reciprocal space integration is performed over
16$\times$16$\times$16 and 10$\times$10$\times$1 \emph{k}-point
grids in the Brillouin zone (BZ) for bulk and thin films,
respectively. The structural parameters of the bulk structure are
fully optimized until the Hellmann-Feynman force on each atom is
less than 0.01 eV/\AA. To model thin films, we use a slab supercell
configuration with a 15 \AA\, thick vacuum layer in the (001)
direction to eliminate the interslab interaction. The free-standing
slab models are constructed with the bulk structural parameters
optimized with optB86b-vdW.

\section{Results}

\subsection{Bulk properties}

\begin{table}[ptb]
\caption{Optimized structural parameters for bulk $\beta$--Bi$_{2}$Pd}.%
\begin{ruledtabular}
\begin{tabular}{lcccccccccccc}
&\multicolumn{3}{c}{Structural parameters}\\
\cline{2-4}
Method&\emph{a} ({\AA})&\emph{c} ({\AA})&$z_{Bi}$\\
\hline
PBE&3.4159&13.1001&0.3644\\
PBE-vdW&3.3904&12.8840&0.3629\\
PBE+SOI&3.4404&13.2507&0.3674\\
PBE$^{\emph{a}}$&3.4140&13.0355&0.3616\\
PBE+SOI$^{\emph{a}}$&3.4060&13.0115&0.3633\\
Expt.$^{\emph{b}}$&3.362&12.983&0.363\\
Expt.$^{\emph{c}}$&3.37&12.96\\
\end{tabular}
$^{\emph{a}}$ Reference \cite{Shein}, $^{\emph{b}}$ Reference \cite{Zhuravlev},
$^{\emph{c}}$ Reference \cite{Imai} \label{lattice}
\end{ruledtabular}
\end{table}

\begin{figure*}[ptb]
\begin{center}
\includegraphics[width=1.0\linewidth]{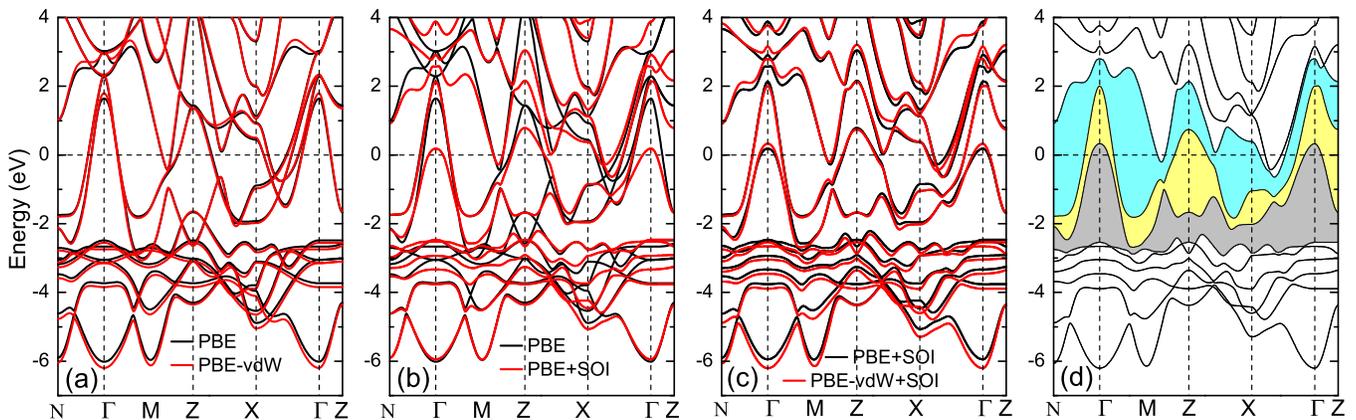}
\end{center}
\caption{(Color online) Bulk band structure of $\beta$--Bi$_{2}$Pd
calculated with PBE, PBE-vdW, PBE+SOI, and PBE-vdW+SOI. Comparison
between (a) PBE (black lines) and PBE-vdW (red lines), (b) PBE
(black lines) and PBE+SOI (red lines), and (c) PBE+SOI (black lines)
and PBE-vdW+SOI (red lines). (d) Band structure calculated with
PBE-vdW+SOI with the band gaps opened due to the SOI highlighted in
cyan, yellow, and gray. The Fermi level is set at zero.
}%
\label{fig2}%
\end{figure*}

\begin{figure*}[ptb]
\begin{center}
\includegraphics[width=1.0\linewidth]{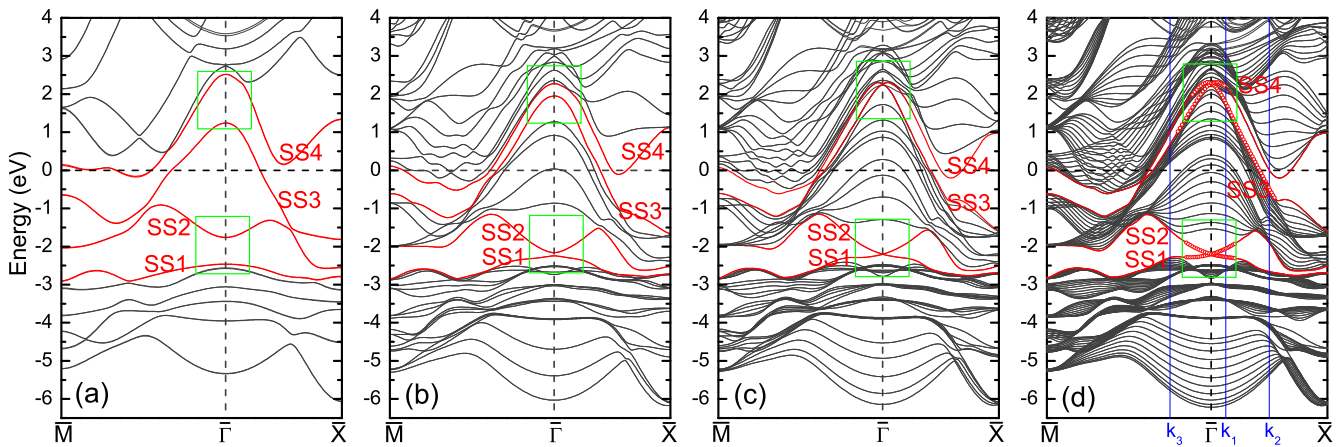}
\end{center}
\caption{(Color online) Band dispersion of $\beta$--Bi$_{2}$Pd thin films for (a) 1,
(b) 3, (c) 5, and (d) 11
 TLs along the \emph{c}-axis. Here, the SS forming the Dirac cone
 are labeled as SS1 and SS2 while the states that form the Rashba splitting as SS3 and SS4.
  The Fermi level is set at zero.}%
\label{fig3}%
\end{figure*}

We begin with a review of the properties of bulk $\beta$--Bi$_{2}$Pd
which crystallizes in the centrosymmetric body-centered tetragonal
(bct) crystal structure with space group \emph{I}4/\emph{mmm}
(No.139) as shown in Figs.~\ref{fig1}(a)-(b). The Wyckoff positions
for the Bi and Pd atoms are 4\emph{e}(0,0,$z_{Bi}$) and
2\emph{a}(0,0,0), respectively. In the conventional unit cell, the
atomic structure can be visualized as a superposition of triple
layers (TLs) with a Bi--Pd--Bi sequence at the center of the unit
cell along the $c$-axis. Within each triple layer every Pd atom
occupies the center of the cube formed by 8 Bi atoms. The bulk BZ
and the BZ projected onto the (001) surface are also shown in
Fig.~\ref{fig1}(d).

In Table I, we list our calculated equilibrium structural parameters
found with PBE, PBE-vdW, and PBE+SOI, together with the
corresponding values from previous theoretical \cite{Shein} and
experimental \cite{Zhuravlev,Imai} studies. PBE-vdW and PBE+SOI
refer to calculations performed with vdW and SOI, respectively.
Overall, our relaxed lattice parameters are in good agreement with
previous calculations performed with the full-potential augmented
plane wave method (FLAPW)-PBE/PBE+SOI \cite{Shein}. When compared to
the experimental values, the in-plane $a$ and out-of-plane $c$
lattice parameters are overestimated by approximately 1.4-1.6\%
(2.1-2.4\%) and 0.9-1.1\% (2.0-2.2\%) in the case of PBE (PBE+SOI).
The inclusion of the vdW dispersion corrections gives a much better
agreement with the experimental values and, in this case, the two
lattice parameters $a$ and $c$ are overestimated and underestimated
by approximately 0.7\%, respectively. The importance of the
long-range dispersive interactions in predicting ground states and
describing the interlayer distance in different classes of materials
has long been
recognized~\cite{Kolmogorov,Mishra,SaPRL,WangSb2Te3,SaNanoscale}.

A simple picture of the chemical bonding can be obtained from
the analysis of the charge density in the (110) plane of $\beta$--Bi$_{2}$Pd.
As shown in Fig.~\ref{fig1}(c), the bond strength distribution predicts strong
covalent bonding within the triple layer (0.035 $e$/au$^{3}$ at the middle of
Bi--Pd bonds) and weak van der Waals interaction between adjacent TLs
(0.012 $e$/au$^{3}$ at the middle of Bi--Bi bonds). These values are similar
to the ones reported in the three-dimensional (3D) TIs Bi$_{2}$Se$_{3}$, Bi$_{2}$Se$_{3}$ and Sb$_{2}$Te$_{3}$
\cite{Mishra,WangBi2Se3,WangSb2Te3}.

To examine the effect of the vdW and SOI on the electronic structure
of bulk $\beta$--Bi$_{2}$Pd, we compare in
Figs.~\ref{fig2}(a)--(d) the band structures obtained
with PBE, PBE-vdW, PBE+SOI, and PBE-vdW+SOI~\cite{PBE-vdW+SOI}. As shown in Fig.~\ref{fig2}(a), the vdW interaction has a limited effect, the bands at the Fermi level remain practically unchanged, while the deeper bands are only
slightly pushed down. The picture is completely different when the
SOI is included, the band structure undergoes significant changes
over the whole BZ [see Fig.~\ref{fig2}(b)]. First, one band at the $\Gamma$ point is downshifted by 1.46 eV and resides just above the Fermi level. Second, a 0.46~eV gap opens at the band crossings around the $Z$ point approximately 2~eV below the
Fermi level [within the gray shaded area in Fig.~\ref{fig2}(d)].
Third, two continuous bulk band gaps are formed across the whole BZ at the Fermi level. These gaps are highlighted in cyan and yellow shades in Fig.~\ref{fig2}(d). Finally, similar to the results without SOI, we find that the reduction in the interlayer spacing due to vdW corrections has only a small effect on the band structure obtained in the presence of SOI [see Fig.~\ref{fig2}(c)].

The inversion symmetry of $\beta$--Bi$_{2}$Pd allows one to perform
a parity analysis and estimate the Z$_2$ invariant for the three SOI-induced
gaps~\cite{Fu08}. By checking the parity at the eight time-reversal
invariant momenta (1$\Gamma$, 2X, 4N, and 1Z), we find that the
middle gap is trivial, while the upper and lower gaps have a
non-trivial topological character in agreement with previous
calculations~\cite{Sakano}. The two topological gaps could harbor
gapless surface states and ARPES measurements have indeed revealed that a Dirac-type band dispersion appears in the lower gap, 2.41~eV below
the Fermi level~\cite{Sakano}. The absence of a gapless surface
state in the upper gap in the ARPES spectra is due to its
localization above the Fermi level. However, a spin polarization
analysis has clearly showed that both topological and trivial
surface states cross the Fermi level~\cite{Sakano}.

The above results demonstrate that while the vdW
dispersion corrections are important for obtaining good
agreement with the experimental reference values for
structural parameters, the SOI is critical for properly
describing the appearance of topological electronic states.

\begin{table}[ptb]
\caption{The band gap at DP and RP for 1-15 $\beta$--Bi$_{2}$Pd TLs calculated with inclusion of SOI.}%
\begin{ruledtabular}
\begin{tabular}{lccccccccccccccccccc}
Slabs&Gap at DP (eV)&Gap at RP (eV)\\
\hline
1 TL&0.703&1.273\\
3 TLs&0.084&0.320\\
5 TLs&0.012&0.000\\
7 TLs&0.002&0.000\\
9 TLs&0.000&0.000\\
11 TLs&0.000&0.000\\
13 TLs&0.000&0.000\\
15 TLs&0.000&0.000\\
\end{tabular}
\label{parityT}
\end{ruledtabular}
\end{table}

\begin{figure}[ptb]
\begin{center}
\includegraphics[width=1.0\linewidth]{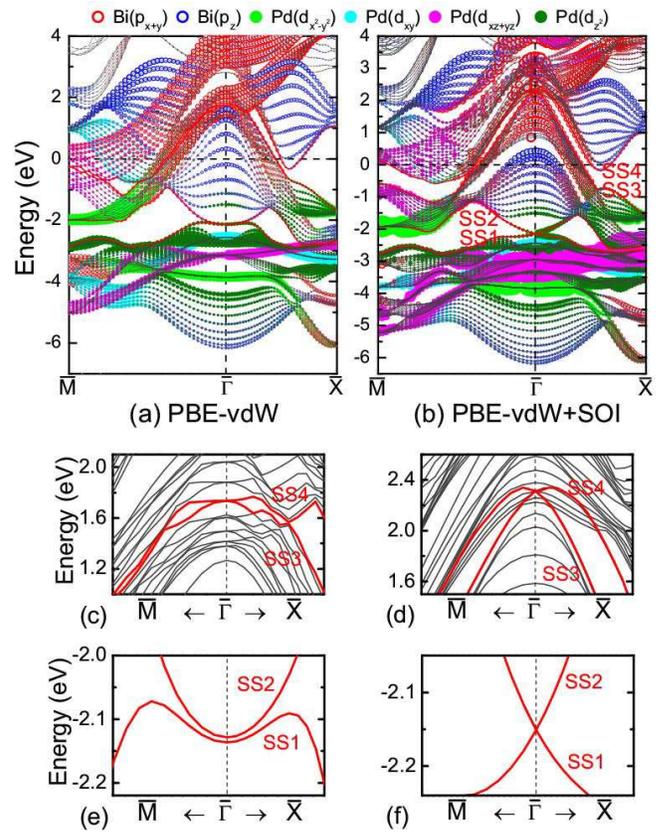}
\end{center}
\caption{(Color online) Projected band dispersion for 11
 TLs calculated with (a) PBE-vdW and (b) PBE-vdW+SOI. The magnified dispersions near the RP and DC are presented in (c) and (e) for PBE-vdW, and (d) and (f) for PBE-vdW+SOI. The Fermi level is set at zero.}%
\label{fig4}%
\end{figure}

\begin{figure}[ptb]
\begin{center}
\includegraphics[width=1\linewidth]{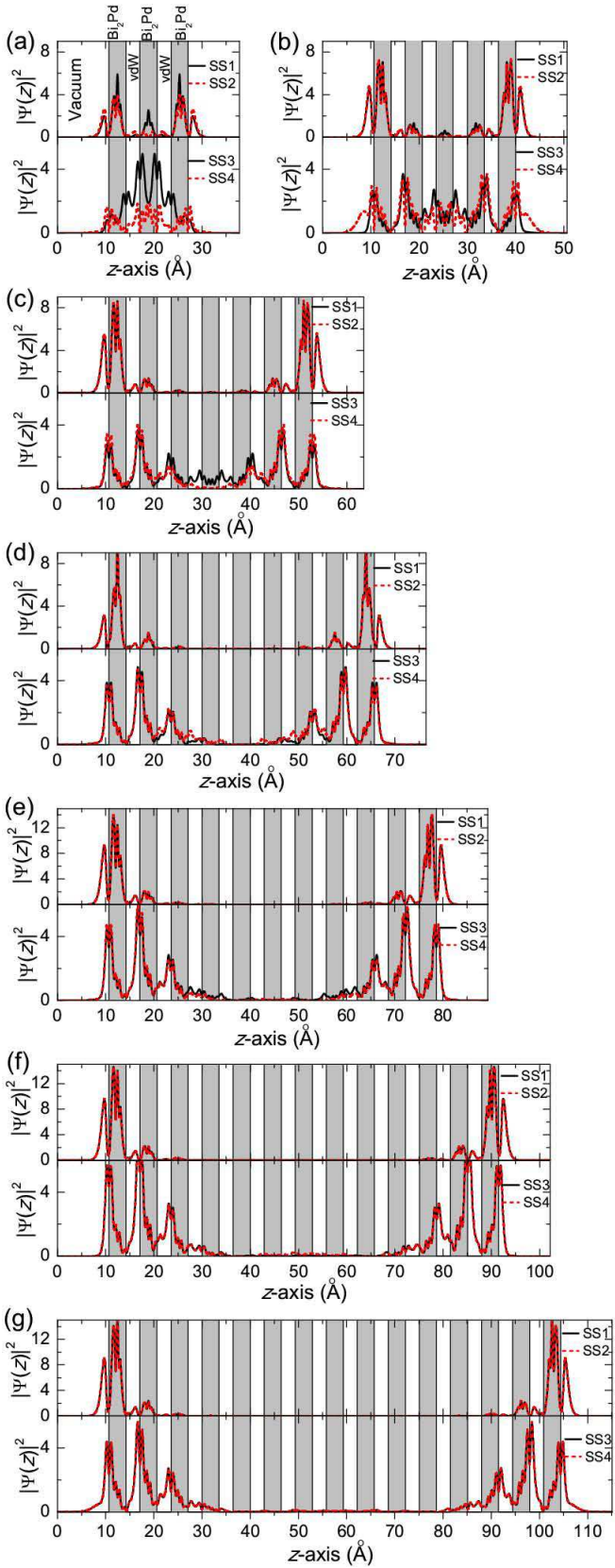}
\end{center}
\caption{(Color online) Spatial distribution of the SS charge density for $\overline{\Gamma}$ point as integrated over (\emph{x},\emph{y}) plane for (a) 3, (b) 5, (c) 7 , (d) 9, (e) 11, (f) 13, and (g) 15 TLs.}%
\label{fig5}%
\end{figure}

\begin{figure}[ptb]
\begin{center}
\includegraphics[width=0.8\linewidth]{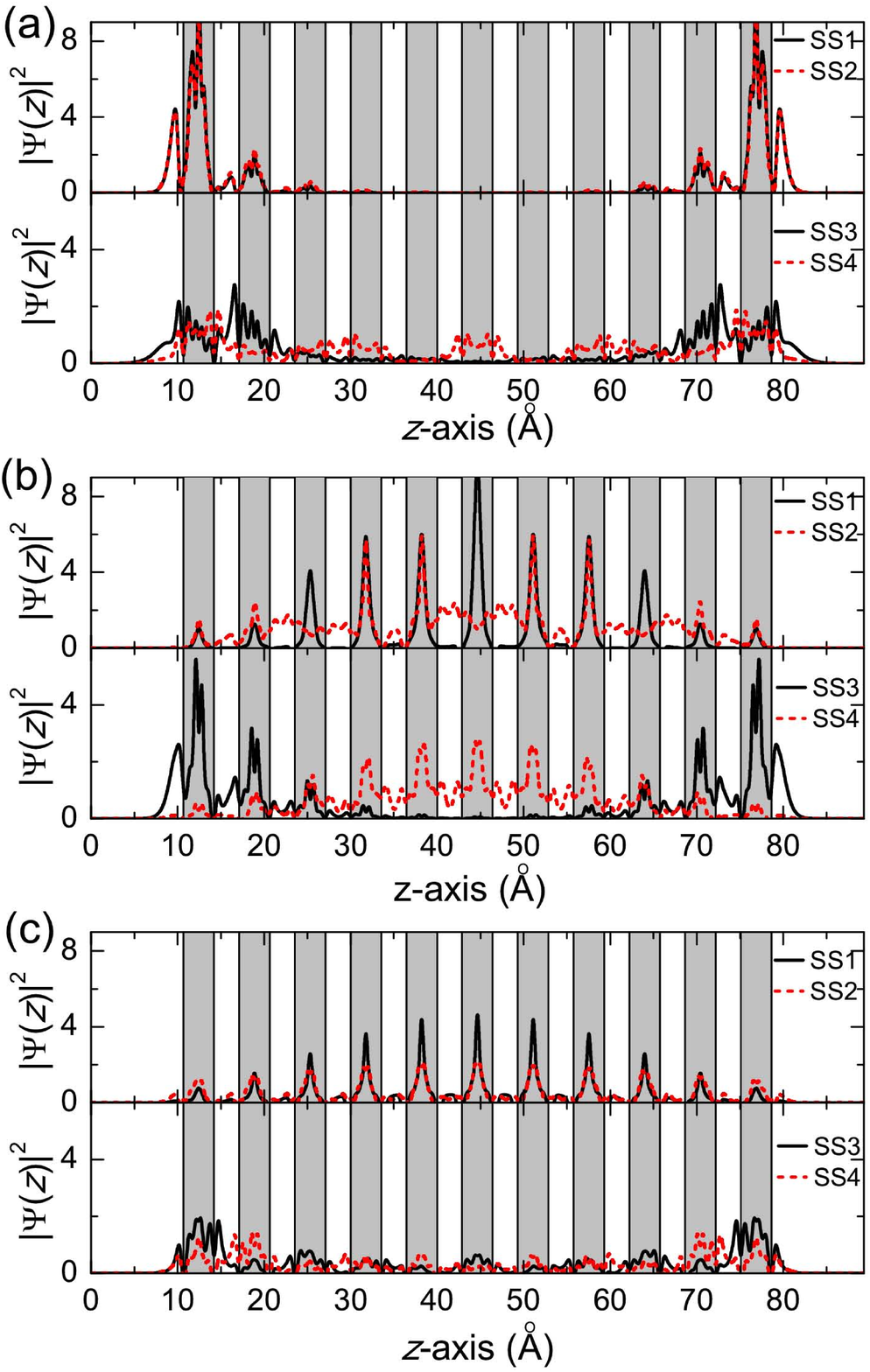}
\end{center}
\caption{(Color online) Spatial distribution of the SS charge density as integrated over (\emph{x},\emph{y}) plane for 11 TLs at (a) $k_{1}$, (b) $k_{2}$, (c) $k_{3}$, and (d) $k_{4}$ points shown in Fig. 3(d).}%
\label{fig6}%
\end{figure}

\subsection{Thin films properties}
We further study the thickness-dependent band structures of
$\beta$--Bi$_{2}$Pd thin films in order to establish the critical
(minimum) thickness at which the topologically protected surface
states develop. We consider several thin films made of 1, 3,
$\cdots$, 15 TLs. Thin films with an even number of TL are excluded
since they break the inversion-symmetry present in the bulk. All
these slab models are constructed with the bulk structural
parameters optimized with PBE-vdW. To check the effect of the
lattice relaxation, we have fully optimized the 3 TLs slab with
PBE-vdW. No visible changes were found between the band structures
calculated with and without SOI in the case of 3 TLs slabs with
optimized and unoptimized lattice parameters.

Figure~\ref{fig3} illustrates the evolution of the surface band
dispersion as a function of slab thickness along the
$\overline{\rm{M}}$-$\overline{\Gamma}$-$\overline{\rm{X}}$
direction. When the film thickness is small enough, the interactions
between the states localized on the top and bottom surfaces lead to
the opening of a gap at the Dirac point (DP). This area is indicated
by the green rectangle in Fig.~\ref{fig3}. Here, we label the
surface states (SSs) that form the Dirac cone as SS1 and SS2. For
1~TL, a relatively large gap of about 0.70~eV is formed -2.45~eV
below the Fermi level at the $\overline{\Gamma}$ point between a
nearly parabolic band (SS2) and an almost flat band (SS1). For
3~TLs, the upper band moves down by 0.40~eV and the lower band moves
up by 0.21~eV, such that the gap is reduced to 0.08~eV. For 5 TLs,
the two bands are almost touching each other at an energy level of
-2.19~eV and the gap is around 0.01~eV. Once the thickness of the
slab reaches 9~TLs, the gap completely disappears and the two bands
with nearly linear dispersion cross each other forming a surface
Dirac cone (S-DC)~\cite{Sakano}. We find that the energy of the DP
is \emph{E$_{\rm{D}}$}=2.20~eV, very close to the
\emph{E$_{\rm{D}}$}=2.41~eV value extracted from the ARPES data
\cite{Sakano}. In Table~\ref{parityT}, we list the gap values at the
DP for all slabs considered. The thickness-dependent behavior of the
gap at the DP found here is similar to the behavior observed in
typical 3D topological
insulators~\cite{He,Li_AdvMat10,WangG,Jiang,Bian12,Yan14}.

Besides the S-DC, an additional topological feature develops in
thicker films at the $\Gamma$ point approximately 2.26~eV above the
Fermi level. As shown in Fig.~\ref{fig3}, with increasing thickness
more bands emerge and two evolve into a Rashba-like crossing point
above 3~TLs. These surface states are labeled as SS3 and SS4 in Fig.
3. The gap values at the Rashba point (RP), presented in
Table~\ref{parityT}, also support this feature. Our finding is
consistent with a previous result reported for a slab of
11~TLs~\cite{Sakano}. We also show in Fig.~\ref{fig4} the
orbital-resolved band dispersion without and with SOI for 11 TLs. At
the DP, both SS1 and SS2 have Pd $d_{z^{2}}$ character, while at the
RP, both SS3 and SS4 have Bi $p_{x+y}$ character. The SOI-induced
crossing at the DP and at the RP can be seen more clearly in
Figs.~\ref{fig4}(c-f).

In order to characterize the spatial spread of the SSs (i.e., the
penetration depth of the surface-state wave functions into the
bulk), we calculate the real space charge distribution of the SSs at
the $\overline{\Gamma}$ point. In Fig.~\ref{fig5}, we show the
charge density integrated over the $xy$--plane as a function of the
$z$ coordinate. For the SS1 and SS2 states, starting from the 7~TL
film, the charge density is concentrated mostly inside the top and
bottom TLs and the wave functions penetrate only 15~\AA\, in depth
from the surface. In contrast, the SS3 and SS4 states only become
surface states starting from the 11~TLs film and have a much larger
penetration depth of about 30~\AA. Based on both the band structure
and the charge density analysis, the critical thickness for
emergence of the Dirac-cone SSs is 9 TLs, while that for the
Rashba-type SSs is 11 TLs.

To further characterize the distribution of the surface states over
atomic layers, we show in Fig.~\ref{fig6} the charge density
integrated over the $xy$--plane for 11 TLs at several $k$-points
in the vicinity of the $\overline{\Gamma}$ point [$k_{1}$--$k_{3}$ points indicated with blue vertical lines in Fig.~\ref{fig3}(d)]. For the $k_{1}$ point closer to $\overline{\Gamma}$, the distribution of the wave functions corresponding to SS1, SS2, and SS3 remains localized to the surface region. On the contrary, the real-space distribution of SS4 displays some weight at the center of slab. Moving away from the $\overline{\Gamma}$ point, the surface states penetrate deeper into the slab and the wave functions acquire a bulk-like spatial distribution. Note that the SS3 still has the characteristic of surface state even at $k_{2}$. Using this charge distribution
analysis, we determine the states away from the $\overline{\Gamma}$
point that retain a charge distribution localized in the surface
region. These are shown as red circles in Fig.~\ref{fig3}(d). The localization of the SSs found in our study is similar to the experimental results for $\alpha$--BiPd~\cite{Benia} and theoretical results for Bi$_2$Se$_3$~\cite{ZhangNJP}.

\section{Conclusion}
In summary, we have investigated the structural, electronic, and
topological properties of the centrosymmetric superconductor
$\beta$--Bi$_{2}$Pd in the bulk and in thin films using
first-principles calculations. We have shown that the structural
parameters obtained with the vdW interaction are in very good
agreement with experiment, and that the inclusion of SOI is
necessary for correctly describing the band structure. In addition,
we have uncovered that thicknesses of 9 and 11 TLs are
required for the appearance of the Dirac and Rashba surface states,
respectively. The penetration depth and the charge distribution of
these states have also been discussed. These findings may prove
important for further exploration of the superconducting state in
$\beta$--Bi$_{2}$Pd thin films.

\section{Acknowledgments}
B. Wang thanks B. Sa and H. Shi for helpful discussions.


\begin{thebibliography}{99}

\bibitem {Fu07}
L. Fu, C. L. Kane, and E. J. Mele, Phys. Rev. Lett. \textbf{98}, 106803 (2007).

\bibitem {Hsieh08}
D. Hsieh, D. Qian, L. Wray, Y. Xia, Y. S. Hor, R. J. Cava, and M. Z. Hasan, Nature \textbf{452}, 970-974 (2008).

\bibitem {ZhangHJ}
H. J. Zhang, C. X. Liu, X. L. Qi, X. Dai, Z. Fang, and S. C. Zhang,
Nat. Phys. \textbf{5}, 438 (2009).

\bibitem {XiaY}
Y. Xia, D. Qian, D. Hsieh, L. Wray, A. Pal, H. Lin, A. Bansil, D. Grauer, Y. S. Hor, R. J. Cava, and M. Z. Hasan, Nat. Phys. \textbf{5}, 398 (2009).

\bibitem {Hsieh09}
D. Hsieh, Y. Xia, D. Qian, L. Wray, J. H. Dil, F. Meier, J. Osterwalder, L. Patthey, J. G. Checkelsky, N. P. Ong, A. V. Fedorov, H. Lin, A. Bansil, D. Grauer, Y. S. Hor, R. J. Cava, and M. Z. Hasan, Nature \textbf{460}, 1101-1105 (2009).

\bibitem {Chen09}
Y. L. Chen, J. G. Analytis, J. H. Chu, Z. K. Liu, S. K. Mo, X. L. Qi, H. J. Zhang, D. H. Lu, X. Dai, Z. Fang, S. C. Zhang, I. R. Fisher, Z. Hussain, and Z. X. Shen, Science \textbf{325}, 178-181 (2009).

\bibitem {Liu}
C. X. Liu, H. J. Zhang, B. Yan, X. L. Qi, T. Frauenheim, X. Dai, Z. Fang, and S. C. Zhang, Phys. Rev. B \textbf{81}, 041307(R) (2010).

\bibitem{Sawai10}
W. Al-Sawai, H. Lin, R. S. Markiewicz, L. A. Wray, Y. Xia, S.-Y. Xu, M. Z. Hasan, and A. Bansil
Phys. Rev. B \textbf{82}, 125208 (2010).

\bibitem{Eremeev10}
S. V. Eremeev, Y. M. Koroteev, and E. V. Chulkov, JETP Lett. \textbf{91}, 594 (2010).

\bibitem{Yang}
K. Yang, W. Setyawan, S. Wang, M. Buongiorno Nardelli, and S. Curtarolo, Nat. Mater. \textbf{11}, 614 (2012).

\bibitem{Kolmogorov16}
S. M. Young, S. Manni, J. Shao, P. C. Canfield, A. N. Kolmogorov,
arXiv:1607.05234v2 (2016).

\bibitem {Hasan10}
M. Z. Hasan and C. L. Kane, Rev. Mod. Phys. \textbf{82}, 3045 (2010).

\bibitem{Bansil16}
A. Bansil, H. Lin, and T. Das, Rev. Mod. Phys. \textbf{88}, 021004 (2016).

\bibitem {XLQi2}X. L. Qi and S. C. Zhang, Rev. Mod. Phys.
\textbf{83}, 1057 (2011).

\bibitem{Hasan15}
M. Z. Hasan, S.-Y. Xu, and G. Bian, Phys. Scr. \textbf{T164}, 014001 (2015).

\bibitem{Sato16}
M. Sato and Y. Ando, arXiv:1608.03395v1 (2016).

\bibitem{Huang_NCOM15}
S.-M. Huang, S.-Y. Xu, I. Belopolski, C.-C. Lee, G. Chang, B. Wang,
N. Alidoust, G. Bian, M. Neupane, C. Zhang, S. Jia, A. Bansil, H.
Lin, and M. Z. Hasan, Nat. Commun. \textbf{6}, 7373 (2015).

\bibitem{Weng_PRX15}
H. Weng, C. Fang, Z. Fang, B. A. Bernevig, and X. Dai Phys. Rev. X
\textbf{5}, 011029 (2015).

\bibitem{Lv_NPHYS15}
B. Q. Lv, N. Xu, H. M. Weng, J. Z. Ma, P. Richard, X. C. Huang, L.
X. Zhao, G. F. Chen, C. E. Matt, F. Bisti, V. N. Strocov, J, Mesot,
Z. Fang, X. Dai, T. Qian, M. Shi, and H. Ding, Nat. Phys.
\textbf{11}, 724 (2015).

\bibitem{Yang_NPHYS15}
L. X. Yang, Z. K. Liu, Y. Sun, H. Peng, H. F. Yang, Y. Zhang, B.
Zhou, Y. Zhang, Y. F. Guo, M. Rahn, D. Prabhakaran, Z. Hussain,
S.-K. Mo, C. Felser, B. Yan, and Y. L. Chen, Nat. Phys. \textbf{11},
728 (2015).

\bibitem{Xu_NPHYS15}
S.-Y. Xu, N. Alidoust, I. Belopolski, Z. Yuan, G. Bian, T.-R. Chang,
H. Zheng, V. N. Strocov, D. Sanchez, G. Chang, C. Zhang, D. Mou, Y.
Wu, L. Huang, C.-C. Lee, S.-M. Huang, B. Wang, A. Bansil, H.-T.
Jeng, T. Neupert, A. Kaminski, H. Lin, S. Jia, and M. Z. Hasan, Nat.
Phys. \textbf{11}, 748 (2015).

\bibitem{Wang_PRB12}
Z. Wang, Y. Sun, X.-Q. Chen, C. Franchini, G. Xu, H. Weng,
X. Dai, and Z. Fang, Phys. Rev. B \textbf{85}, 195320 (2012).

\bibitem{Ali14}
M. N. Ali, Q. Gibson, S. Jeon, B. B. Zhou, A. Yazdani, and R. J.
Cava, Inorg. Chem. \textbf{53}, 4062 (2014).

\bibitem{Borisenko_PRL14}
S. Borisenko, Q. Gibson, D. Evtushinsky, V. Zabolotnyy, B.
B\"{u}chner, and R. J. Cava,  Phys. Rev. Lett. \textbf{113}, 027603
(2014).

\bibitem{Liu_SCI14}
Z. K. Liu, B. Zhou, Y. Zhang, Z. J. Wang, H. M. Weng, D.
Prabhakaran, S.-K. Mo, Z. X. Shen, Z. Fang, X. Dai, Z. Hussain, and
Y. L. Chen, Science \textbf{343}, 864 (2014).

\bibitem{Neupane_NCOM14}
M. Neupane, S.-Y. Xu, R. Sankar, N. Alidoust, G. Bian, C. Liu, I.
Belopolski, T.-R. Chang, H.-T. Jeng, H. Lin, A. Bansil, F. Chou, and
M. Z. Hasan, Nat. Commun. \textbf{5}, 3786 (2014).

\bibitem{Bian_NCOM16}
G. Bian, T.-R. Chang, R. Sankar, S.-Y. Xu, H. Zheng, T. Neupert,
C.-K. Chiu, S.-M. Huang, G. Chang, I. Belopolski, D. S. Sanchez, M.
Neupane, N. Alidoust, C. Liu, B. Wang, C.-C. Lee, H.-T. Jeng, C.
Zhang, Z. Yuan, S. Jia, A. Bansil, F. Chou, H. Lin, and M. Z. Hasan,
Nat. Commun. \textbf{7}, 10556 (2016).

\bibitem{Wu_NPHYS16}
Y. Wu, L.-L. Wang, E. Mun, D. D. Johnson, D. Mou, L. Huang, Y. Lee,
S. L. Bud$'$ko, P. C. Canfield, and A. Kaminski, Nat. Phys.
\textbf{12}, 667 (2016).

\bibitem{Schnyder-PRB15}
A. P. Schnyder, S. Ryu, A. Furusaki, and A. W. W. Ludwig, Phys. Rev. B \textbf{78},
195125 (2008).

\bibitem{Ando15}
Y. Ando and L. Fu, Annu. Rev. Condens. Matter Phys. \textbf{6},
361-381 (2015).

\bibitem{Xu15}
J. P. Xu, M. X. Wang, Z. L. Liu, J. F. Ge, X. Yang, C. Liu, Z. A. Xu, D. Guan, C. L. Gao, D. Qian, Y. Liu, Q. H. Wang, F. C. Zhang, Q. K. Xue, and J. F. Jia, Phys. Rev. Lett. \textbf{114}, 017001 (2015).

\bibitem{Fu08}
L. Fu and C. L. Kane, Phys. Rev. Lett. \textbf{100}, 096407 (2008).

\bibitem {Nayak}
C. Nayak, S. H. Simon, A. Stern, M. Freedman, and
S. D. Sarma, Rev. Mod. Phys. \textbf{80}, 1083 (2008).

\bibitem{Fu10}
L. Fu and E. Berg, Phys. Rev. Lett. \textbf{105}, 097001 (2010).

\bibitem{Hor}
Y. S. Hor, A. J. Williams, J. G. Checkelsky, P.
Roushan, J. Seo, Q. Xu, H. W. Zandbergen, A. Yazdani, N. P. Ong, and
R. J. Cava, Phys. Rev. Lett. \textbf{104}, 057001 (2010).

\bibitem{Kriener}
M. Kriener, K. Segawa, Z. Ren, S. Sasaki, and Y. Ando, Phys. Rev. Lett. \textbf{106}, 127004 (2011).

\bibitem{Sasaki11}
S. Sasaki, M. Kriener, K. Segawa, K. Yada, Y. Tanaka, M. Sato, and Y. Ando, Phys. Rev. Lett. \textbf{107}, 217001 (2011).

\bibitem{Sasaki12}
S. Sasaki, Z. Ren, A. A. Taskin, K. Segawa, L. Fu, and Y. Ando,
Phys. Rev. Lett. \textbf{109}, 217004 (2012).

\bibitem{ZhangPNAS}
J. L. Zhang, S. J. Zhang, H. M. Weng, W. Zhang,
L. X. Yang, Q. Q. Liu, S. M. Feng, X. C. Wang, R. C. Yu, L. Z. Cao,
L. Wang, W. G. Yang, H. Z. Liu, W. Y. Zhao, S. C. Zhang, X. Dai, Z.
Fang, and C. Q. Jin, \emph{Proc. Nat. Acad. Sci.} \textbf{108}, 24
(2011).

\bibitem{ZhuSR}
J. Zhu, J. L. Zhang, P. P. Kong, S. L. Zhang, X. H. Yu, J. L. Zhu, Q. Q. Liu, X. Li, R. C. Yu, R. Ahuja, W. G. Yang, G.
Y. Shen, H. K. Mao, H. M. Weng, X. Dai, Z. Fang, Y. S. Zhao, and C.
Q. Jin, Sci. Rep. \textbf{3}, 2016 (2013).

\bibitem{Kirshenbaum-PRL}
K. Kirshenbaum, P. S. Syers, A. P. Hope,
N. P. Butch, J. R. Jeffries,
 S. T. Weir, J. J. Hamlin, M. B. Maple, Y. K. Vohra, and J. Paglione,
\textbf{111}, 087001 (2013).

\bibitem{KongPP}
P. P. Kong, J. L. Zhang, S. J. Zhang, J. Zhu, Q.
Q. Liu, R. C. Yu, Z. Fang, C. Q. Jin, W. G. Yang, X. H. Yu, J. L.
Zhu, and Y. S. Zhao, J. Phys.: Condens. Matter \textbf{25}, 362204
(2013).

\bibitem{Mondal}
M. Mondal, B. Joshi, S. Kumar, A. Kamlapure, S.
C. Ganguli, A. Thamizhavel, S. S. Mandal, S. Ramakrishnan, and P.
Raychaudhuri, Phys. Rev. B \textbf{86}, 094520 (2012).

\bibitem{Sun}
Z. Sun, M. Enayat, A. Maldonado, C. Lithgow, E.
Yelland, D. C. Peets, A. Yaresko, A. P. Schnyder, and P. Wahl, Nat.
Comm. \textbf{6}, 6633 (2015).

\bibitem{Neupane}
M. Neupane, N. Alidoust, M. M. Hosen, J.-X Zhu, K. Dimitri, S.-Y.
Xu, N. Dhakal, R. Sankar, I. Belopolski, D. S. Sanchez, T. R. Chang,
H.-T. Jeng, K. Miyamoto, T. Okuda, H. Lin, A. Bansil, D.
Kaczorowski, F. Chou, M. Z. Hasan, and T. Durakiewicz, Nat. Comm.
\textbf{7}, 13315 (2015).

\bibitem{Sakano}
M. Sakano, K. Okawa, M. Kanou, H. Sanjo, T.
Okuda, T. Sasagawa, and K. Ishizaka, Nat. Comm. \textbf{6}, 8595
(2015).

\bibitem{Herrera}
E. Herrera, I. Guillam\'{o}n, J. A. Galvis, A. Correa, A. Fente, R.
F. Luccas, F. J. Mompean, M. Garc\'{\i}a-Hern\'{a}ndez, S. Vieira,
J. P. Brison et al., Phys. Rev. B \textbf{92}, 054507 (2015).

\bibitem{Che}
L. Che, T. Le, C. Q. Xu, X. Z. Xing, Z. Shi, X. Xu, and X. Lu, Phys.
Rev. B \textbf{94}, 024519 (2016).

\bibitem{Kacmarcik}
J. Ka\u{c}mar\u{c}\'{i}k, Z. Pribulov\'{a}, T. Samuely, P.
Szab\'{o}, V. Cambel, J. \u{S}olt\'{y}s, E. Herrera, H. Suderow, A.
Correa-Orellana, D. Prabhakaran, and P. Samuely, Phys. Rev. B {\bf
93}, 144502 (2016).

\bibitem{Biswas}
P. K. Biswas, D. G. Mazzone, R. Sibille, E. Pomjakushina, K. Conder,
H. Luetkens, C. Baines, J. L. Gavilano, M. Kenzelmann, A. Amato, and
E. Morenzoni, Phys. Rev. B {\bf 93}, 220504 (2016).

\bibitem{Margine}
J.-J. Zheng and E. R. Margine, under review.

\bibitem{LvYF}
Y. F. Lv, W. L. Wang, Y. M. Zhang, H. Ding, W. Li,
L. Wang, K. He, C. L. Song, X. C. Ma, and Q. K. Xue,
arXiv:1607.07551 (2016).

\bibitem{Ali_PRB14}
M. N. Ali, Q. D. Gibson, T. Klimczuk, and R. J. Cava, Phys. Rev. B {\bf 89}, 020505(R) (2014).

\bibitem{Chang_PRB16}
T. R. Chang, P. J. Chen, G. Bian, S. M. Huang, H. Zheng, T. Neupert,
R. Sankar, S. Y. Xu, I. Belopolski, G. Chang, B. K. Wang, F. Chou,
A. Bansil, H. T. Jeng, H. Lin, and M. Z. Hasan, Phys. Rev. B {\bf
93}, 245130 (2016).

\bibitem{He}
Y. Zhang, K. He, C. Z. Chang, C. L. Song, L. L. Wang,
X. Chen, J. F. Jia, Z. Fang, X. Dai, W. Y. Shan, S. Q. Shen, Q. Niu,
X. L. Qi, S. C. Zhang, X. C. Ma, and Q. K. Xue, Nat. Phys.
\textbf{6}, 584 (2010).

\bibitem{Li_AdvMat10}
Y. Y. Li, G. Wang, X. G. Zhu, M. H. Liu, C.
Ye, X. Chen, Y. Y. Wang, K. He, L. L. Wang, X. C. Ma, H. J. Zhang,
X. Dai, Z. Fang, Y. Liu, X. L. Qi, J. F. Jia, S. C. Zhang, and Q. K.
Xue, Adv. Mater.  \textbf{22}, 4002 (2010).

\bibitem{WangG}
G. Wang, X. Zhu, J. Wen, X. Chen, K. He, L. Wang,
X. Ma, Y. Liu, X. Dai, Z. Fang, J. Jia, and Q. Xue, Nano Res.
\textbf{3}, 874 (2010).

\bibitem{Kresse3}
G. Kresse and J. Furthm\"{u}ller, Phys. Rev. B
\textbf{54}, 11169 (1996).

\bibitem{PBE}
J. P. Perdew, K. Burke, and M. Ernzerhof, Phys. Rev.
Lett. \textbf{77}, 3865 (1996).

\bibitem{vdW1}
J. Klime\v{s}, D. R. Bowler, and A. Michaelides, J.
Phys.: Condens. Matter \textbf{22}, 022201 (2010).

\bibitem{vdW2}
J. Klime\v{s}, D. R. Bowler, and A. Michaelides,
Phys. Rev. B \textbf{83}, 195131 (2011).

\bibitem{SOI}
D. D. Koelling and B. N. Harmon, J. Phys. C: Solid
State Phys. \textbf{10}, 3107 (1977).

\bibitem{Shein}
I. R. Shein and A. L. Ivanovskii, J. Supercond.
Nov. Magn. \textbf{26}, 1-4 (2013).

\bibitem{Imai}
Y. Imai, F. Nabeshima, T. Yoshinaka, K. Miyatani,
R. Kondo, S. Komiya, I. Tsukada, and A. Maeda, J. Phys. Soc. Jpn.
\textbf{81}, 113708 (2012).

\bibitem {Zhuravlev}
N. N. Zhuravlev, Zh. Eksp. Teor. Fiz.
\textbf{32}, 1305 (1957).

\bibitem{Kolmogorov}
A. Bil, B. Kolb, R. Atkinson, D. G. Pettifor,
T. Thonhauser, and A. N. Kolmogorov, Phys. Rev. B \textbf{83},
224103 (2011).

\bibitem{Mishra}
S. K. Mishra, S. Satpathy, and O. Jepsen, J.
Phys.: Condens. Matter \textbf{9}, 461 (1997).

\bibitem{SaPRL}
B. Sa, J. Zhou, Z. Sun, J. Tominaga, and R. Ahuja, Phys. Rev. Lett. \textbf{109}, 096802 (2012).

\bibitem{WangSb2Te3}
B. T. Wang, P. Souvatzis, O. Eriksson, and P. Zhang, J. Chem. Phys. \textbf{142}, 174702 (2015).

\bibitem{SaNanoscale}
B. Sa, Z. Sun, and B. Wu, Nanoscale \textbf{8}, 1169 (2016).

\bibitem {WangBi2Se3}
B. T. Wang and P. Zhang, Appl. Phys. Lett. \textbf{100}, 082109 (2012).

\bibitem{PBE-vdW+SOI} PBE-vdW+SOI refers to calculations with SOI
using the structural parameters optimized with PBE-vdW.

\bibitem{Jiang}
Y. Jiang, Y. Wang, M. Chen, Z. Li, C. Song, K. He,
L. Wang, X. Chen, X. Ma, and Q. K. Xue, Phys. Rev. Lett.
\textbf{108}, 016401 (2012).

\bibitem{Bian12}
G. Bian, X. Wang, Y. Liu, T. Miller, and T. C.
Chiang, Phys. Rev. Lett. \textbf{108}, 176401 (2012).

\bibitem{Yan14}
C. Yan, J. Liu, Y. Zang, J. Wang, Z. Wang, P. Wang,
Z. D. Zhang, L. Wang, X. Ma, S. Ji, K. He, L. Fu, W. Duan, Q. K.
Xue, and X. Chen, Phys. Rev. Lett. \textbf{112}, 186801 (2014).

\bibitem{Benia}
H. M. Benia, E. Rampi, C. Trainer, C. M. Yim, A.
Maldonado, D. C. Peets, A. St\"{o}hr, U. Starke, K. Kern, A.
Yaresko, G. Levy, A. Damascelli, C. R. Ast, A. P. Schnyder, and P.
Wahl, Phys. Rev. B \textbf{94}, 121407(R) (2016).

\bibitem{ZhangNJP}
W. Zhang, R. Yu, H. J. Zhang, X. Dai, and Z. Fang, New J. Phys.
\textbf{12}, 065013 (2010).

\end{thebibliography}
\end{document}